\documentclass[aps,pre,showpacs,reprint,groupedaddress,amssymb]{revtex4-1}

\usepackage{lmodern}
\usepackage[T1]{fontenc}
\usepackage{color}
\usepackage{amsmath}
\usepackage{graphicx}
\usepackage[dvipsnames]{xcolor}
\usepackage{bm}
\usepackage{epstopdf}

\begin{document}

\title{Optimal waveform for the entrainment of oscillators perturbed by an amplitude-modulated high-frequency force}


\author{Viktor Novi\v{c}enko}
\email[]{novicenko@pfi.lt}
\homepage[]{http://www.itpa.lt/~novicenko/}
\affiliation{Institute of Theoretical Physics and Astronomy, Vilnius University, Saul\.{e}tekio Avenue 3, LT-10222 Vilnius, Lithuania}

\author{Irmantas Ratas}

\affiliation{Center for Physical Sciences and Technology, Saul\.{e}tekio Avenue 3, LT-10222 Vilnius, Lithuania}

\begin{abstract}

We analyze limit cycle oscillators under  perturbation constructed as a product of two signals, namely, an  envelope with a period close to natural period of an oscillator and a high-frequency carrier signal. A theory for obtaining an envelope waveform that achieves the maximal frequency interval of entrained oscillators is presented. The  optimization problem for fixed power and maximal allowed amplitude is solved by employing the phase reduction method and the Pontryagin's maximum principle. We have shown that the optimal envelope waveform is a bang-bang-type solution. Also, we have found ``inversion'' symmetry that relates two signals with different powers, but the same interval of entrained frequencies. The theoretical results are confirmed numerically on FitzHugh-Nagumo oscillators.

\end{abstract}

\pacs{05.45.Xt, 02.30.Xx, 87.19.lr}

\maketitle

\section{\label{sec1} Introduction}

The entrainment phenomenon, when oscillating systems are asymptotically synchronized to an external periodic signal~\cite{pikov01,kura03}, is widely used in many scientific and engineering applications. The ability to optimize entrainment is essential for achieving cardiac resynchronization~\cite{Houthuizen2010}, quick adjustment from jet lag~\cite{Waterhouse2007}, maximizing the growth rate of plants~\cite{McClung2011}, implementing phase-locked loop circuits and injection-locked microintegrated oscillators~\cite{Feng2008}. 

The development of the optimal stimulation waveforms that manage to drive complex systems into the desired conditions is an important challenge met in the neuroscience. For example, the deep brain stimulation is a clinically approved therapeutic procedure for the treatment of Parkinson's disease, essential tremor and dystonia~\cite{benabid1991,marks2005}, where  electrical stimuli are used to suppress pathological synchrony of the neurons~\cite{Lozano2004}. One of the stimulation techniques, called a coordinated reset neuromodulation~\cite{tass2003,tass2012,tass2014}, suppress a mean field of a neural population via amplitude-modulated high-frequency electrical signals, which are periodically delivered at different sites of the population with shifted phases. The efficiency of this technique depends on a number of neurons synchronized with envelope of the electrical signal.

In the past decade numerous theoretical works addressing the waveform optimization problems have been investigated. For example, the optimal current that elicits a neuron to spike at a defined time~\cite{moeh2006}, the minimum power waveform that is capable to entrain oscillators~\cite{harada10}, the input that minimizes the average transient time required to entrain oscillators~\cite{zlotnik2013}, the signals that minimize control energy or the transient time for the subharmonic entrainment of forced oscillators~\cite{Zlotnik2014}, optimization for minimum power of bounded~\cite{dasa11} and charge-balanced~\cite{Dasanayake2015} stimuli for entrainment. An interesting relation between the maximization of the locking range of the oscillators and maximization of the Tsallis entropy was shown in Ref.~\cite{Tanaka2015}. All these works assume that the external force is weak and the well-known phase reduction method~\cite{kura03} can be applied. Nevertheless, in practical problems, like in the deep brain stimulation, the weak force assumption is not always the case. 

Recently, the extension of the phase reduction method for the limit cycle oscillators under a strong amplitude-modulated high-frequency (AMHF) force  was suggested~\cite{Pyragas2015}. An equation for the phase dynamics was derived by combining the conventional phase reduction approach~\cite{kura03} and an averaging method~\cite{sand07,burd07}. In Ref.~\cite{Pyragas2015} the extended phase reduction method was used to derive an optimal waveform of the AMHF perturbation that ensures an entrainment of the oscillator with a minimal power.

Motivated by the requirements met in the application of coordinated reset neuromodulation, we formulate the AMHF envelope optimization problem to attain the maximum frequency interval of the entrained oscillators when the power and maximal allowed amplitude of the stimulation signal is fixed. The main difference between problems formulated in this paper and Ref.~\cite{Pyragas2015} is that here a maximal frequency interval can include both positive and negative mismatches. By employing the extended phase reduction method~\cite{Pyragas2015} and the Pontryagin maximum principle~\cite{pontryagin1962}, we establish analytic conditions for the optimal waveform. It is shown that the optimal envelope contains only the intervals of maximal and zero amplitude of stimulation. Additionally, we have found that any waveform can be ``inverted'' and it gives the same frequency interval of the entrained oscillators but with different power of the perturbation. In the case of the optimal waveform the ``inversion'' symmetry means alteration of maximal amplitude to zero and vice versa.

The paper is organized as follows. Section~\ref{sec_phas_red} is devoted to presenting the phase reduction method extended for the strong AMHF perturbation. The optimization problem is formulated and analyzed in Sec.~\ref{sec_opt}, where also the two analytically tractable cases are examined. In Sec.~\ref{sec_FHN} a numerical confirmation of the theory is demonstrated on the FitzHugh-Nagumo neuron model. A summary is presented in Sec.~\ref{sec_disc}.

\section{\label{sec_phas_red} Phase reduction of limit cycle oscillators under strong amplitude-modulated high-frequency force}

Let us consider a family of uncoupled and unperturbed dynamical systems $\dot{\mathbf{x}}^{(a)}=\mathbf{f}^{(a)}(\mathbf{x}^{(a)})$ with a $n$-dimensional state vector $\mathbf{x}^{(a)}(t) \in \mathbb{R}^n$ of the system where superscript $^{(a)}$ denotes parametric dependence of the state vector, and $\mathbf{f}^{(a)}(\mathbf{x}): \mathbb{R}^n \times \mathbb{R} \rightarrow \mathbb{R}^n$ is a vector field, which represents the free dynamics. All unperturbed systems have a stable periodic solution $\bm{\xi}^{(a)}(t+T^{(a)})=\bm{\xi}^{(a)}(t)$ with the period~$T^{(a)}$. We are interested in the dynamics of these oscillators under the strong AMHF force:
\begin{equation}
\dot{\mathbf{x}}^{(a)}=\mathbf{f}^{(a)}(\mathbf{x}^{(a)})+\mathbf{u}_1 K \psi(\Omega t) \varphi(\omega t),
\label{main}
\end{equation}
where the constant vector $\mathbf{u}_1=(1,0,\hdots,0)^T$ represents an assumption that only the first dynamical variable can be affected and the parameter $K$ is a perturbation amplitude. Functions $\psi(s)$ and $\varphi(s)$ are  $2\pi$-periodic and stands for the slowly varying envelope and high-frequency (HF) carrier signal, respectively. We require that the average of the HF waveform vanishes: ${\left\langle \varphi \right\rangle=(2\pi)^{-1}\int_0^{2\pi} \varphi(s) \mathrm{d}s=0}$. In the terms of neurostimulation, this constraint represents a charge-balanced requirement, which is clinically mandatory to avoid tissue damage~\cite{harnack2004,tass08}. A ratio between the carrier and envelope frequencies $\omega/\Omega$ $(\omega\gg\Omega)$ is assumed to be an integer number so the product $\psi(\Omega t)\varphi(\omega t)$ is also a periodic function with the same period $T=2\pi/\Omega$ as the envelope. In order to uniquely factorize perturbation into $K$, $\psi(s)$, and $\varphi(s)$ parts, we assume that the maximum of the function $\varphi(s)$ is equal to $1$ and the minimum is not below $-1$, moreover the envelope $\psi(s)$ is in the interval $[-1,1]$ and at least one time during the period it reaches one of the boundary.

We are interested in the case when the amplitude $K$ is  comparable  with the corresponding elements of the vector field $\mathbf{f}^{(a)}(\mathbf{x}^{(a)})$, and the high-frequency $\omega \to \infty$. This means that for the system Eq.~(\ref{main}) a conventional phase reduction approach can not be applied. Therefore, we refer to the phase reduction method extended for the oscillators under the strong AMHF perturbation~\cite{Pyragas2015}. Following Ref.~\cite{Pyragas2015}, we replace the set of parameters $(K,\omega)$ by the set of parameters $(A,\omega)$, where $A=K/\omega$. Due to the one-to-one relation between the above parameter spaces, the solution found in the space of the parameters $(A,\omega)$ can be uniquely transformed into the original space of the parameters $(K,\omega)$.  

The phase dynamics of the Eq.~(\ref{main}) reads
\begin{equation}
\dot{\vartheta}^{(a)}=1+\frac{\left\langle \Phi^2 \right\rangle}{2} A^2 z_{\mathrm{eff}}^{(a)}\left(\vartheta^{(a)}\right)\psi^2(\Omega t)+O(A^3),
\label{phase_main}
\end{equation}
where the angle brackets $\left\langle \cdots \right\rangle=(2\pi)^{-1}\int_0^{2\pi}\cdots \mathrm{d}s$ denote the averaging of a function over its period, the function $\Phi(s)$ defined as
\begin{equation}
\Phi(s)=\int_0^{s} \varphi(s_1)\mathrm{d}s_1-(2\pi)^{-1}\int_0^{2\pi} \int_0^{s_1} \varphi(s_2)\mathrm{d}s_2 \mathrm{d}s_1, \label{Phi_def}
\end{equation}
is a particular antiderivative of the HF function $\varphi(s)$, and $z^{(a)}_{\mathrm{eff}}(\vartheta)$ is an effective PRC defined as a dot product of an infinitesimal PRC $\mathbf{z}^{(a)}(\vartheta)$ of the oscillator and a second derivative of the vector flow $\mathbf{f}^{(a)}$ with respect to the first dynamical variable calculated on the limit cycle:
\begin{equation}
z^{(a)}_{\mathrm{eff}}(\vartheta)=\left[\mathbf{z}^{(a)}(\vartheta) \right]^T \cdot\left.\frac{\partial^2 \mathbf{f}^{(a)}(\mathbf{x})}{\partial x_1^2}\right|_{\mathbf{x}=\bm{\xi}^{(a)}(\vartheta)}.
\label{eff_prc}
\end{equation}
From Eq.~(\ref{phase_main}) one can see that the sign of the function $\psi(s)$ does not influence on the phase dynamics, thus we can consider $\psi(s) \in [0,1]$. Also we note that $A^2$ is a small parameter of the phase reduction. Thus all terms smaller than $A^2$ will be neglected.

At this point we have to assume that the parameter ${a \in [a_1,a_2]}$ is such that the natural frequency of the oscillator $\Omega^{(a)}=2\pi/T^{(a)}$ is a monotonic function on $a$ and the value $\Omega^{(a)}$ is close to $\Omega$ in that interval, i.e., $\Omega^{(a)}-\Omega=O(A^2)$. Our goal will be to optimize the envelope $\psi(s)$ in order to attain maximal frequency locking interval. In this context, we can restrict ourselves on the analysis of two boundary oscillators $a\equiv\pm$ with the natural frequencies $\Omega^{(\pm)}$, where $\Omega^{(+)}$ ($\Omega^{(-)}$) is the highest (lowest) frequency that can synchronize with the frequency of the envelope $\Omega$. As we will see later, all oscillators with the frequencies in between of $\Omega^{(-)}$ and $\Omega^{(+)}$ are also synchronized with the envelope.

Let us denote the effective PRC $z^{(a_0)}_{\mathrm{eff}}(\vartheta)\equiv z_{\mathrm{eff}}(\vartheta)$ at the parameter value $a=a_0$, where the oscillator's natural frequency $\Omega^{(a_0)}$ is equal to $\Omega$. The effective PRCs of the boundary oscillators $(\pm)$ are close to $z_{\mathrm{eff}}(\vartheta)$:
\begin{equation}
z^{(\pm)}_{\mathrm{eff}}(s/\Omega^{(\pm)})=z_{\mathrm{eff}}(s/\Omega)+O(A^2).
\label{eff_prc_apr}
\end{equation}
Hence, the phase dynamics of the boundary oscillators reads
\begin{equation}
\dot{\vartheta}^{(\pm)}=1+\frac{\left\langle \Phi^2 \right\rangle}{2} A^2 z_{\mathrm{eff}}\left(\frac{\Omega^{(\pm)}}{\Omega}\vartheta^{(\pm)}\right)\psi^2(\Omega t).
\label{phase_bound}
\end{equation}
We are interested in the difference between the oscillator's and the envelope's phases, therefore we introduce a new phase variables $\phi^{(\pm)}(t)=\Omega^{(\pm)}\vartheta^{(\pm)}(t)-\Omega t$. By changing the time scale $\tau=\Omega t$ and having in mind that $\Omega^{(\pm)}=\Omega+O(A^2)$, Eq.~(\ref{phase_bound}) transforms to
\begin{equation}
\frac{\mathrm{d}\phi^{(\pm)}}{\mathrm{d}\tau}=\left(\frac{\Omega^{(\pm)}}{\Omega}-1 \right)+\frac{\left\langle \Phi^2 \right\rangle}{2} A^2 z_{\mathrm{eff}}\left(\frac{\phi^{(\pm)}+\tau}{\Omega}\right)\psi^2(\tau).
\label{phase_bound1}
\end{equation}
Both terms in the right hand side of Eq.~\eqref{phase_bound1} are of the order of $O(A^2)$, hence for this equation we can apply the averaging method~\cite{sand07,burd07}. The averaged phases $\bar{\phi}^{(\pm)}(\tau)$ satisfy the differential equations
\begin{equation}
\frac{\mathrm{d}\bar{\phi}^{(\pm)}}{\mathrm{d}\tau}=\left(\frac{\Omega^{(\pm)}}{\Omega}-1 \right)+A^2 H\left(\bar{\phi}^{(\pm)} \right),
\label{phase_aver}
\end{equation}
with the $2\pi$-periodic function
\begin{equation}
H(\chi)=\frac{1}{2\pi}\frac{\left\langle \Phi^2 \right\rangle}{2}\int\limits_0^{2\pi} \tilde{z}_{\mathrm{eff}}(\chi+s)\psi^2(s)\mathrm{d}s,
\label{H_def}
\end{equation}
where  $\tilde{z}_{\mathrm{eff}}(s)=z_{\mathrm{eff}}(s/\Omega)$ is rescaled effective PRC. 

The entrainment occurs when the phases $\bar{\phi}^{(\pm)}$ will be locked or, in other words, the differential Eq.~\eqref{phase_aver} will have fixed points. We denote a point $\chi^+$ ($\chi^-$) where the function $H(\chi)$ is maximal (minimal), that is $H(\chi^+)=\mathrm{max}[H(\chi)]$ ($H(\chi^-)=\mathrm{min}[H(\chi)]$). The fixed points of Eqs.~\eqref{phase_aver} are $\bar{\phi}^{(\pm)}_{\mathrm{fix}}=\chi^{\mp}$. Finally, the boundary frequencies can be estimated from
\begin{equation}
\Omega^{(\pm)}=\Omega\left[1-A^2 H(\chi^{\mp})\right].
\label{bound_freq}
\end{equation}
Assuming that $H(\chi)$ is continuous, any oscillator with the frequency $\Omega^{(a)} \in [\Omega^{(-)},\Omega^{(+)}]$ will synchronize with the envelope, since Eq.~(\ref{phase_aver}) for the frequency $\Omega^{(a)}$ will have at least one stable fixed point.

In the envelope's waveform optimization problem, one needs to maximize the frequency locking interval
\begin{equation}
\Delta \Omega = \Omega^{(+)}-\Omega^{(-)}=\Omega A^2\left[ H(\chi^+)-H(\chi^-)\right].
\label{inter}
\end{equation}
The envelope waveform has an interesting symmetry: it can be ``inverted'' and still will have the same frequency locking interval. Let us define an ``inverted'' envelope as $\psi_{\mathrm{inv}}(\tau)=\sqrt{1-\psi^2(\tau)}$. By Eq.~\eqref{H_def}, an ``inverted'' envelope  $\psi_{\mathrm{inv}}(\tau)$ will give a function ${H_{\mathrm{inv}}(\chi)=\frac{\left\langle \Phi^2 \right\rangle}{2}\left\langle \tilde{z}_{\mathrm{eff}} \right\rangle-H(\chi)}$. This  function will have the maximum and minimum at the points $\chi_{\mathrm{inv}}^+=\chi^-$ and $\chi_{\mathrm{inv}}^-=\chi^+$, respectively. Then the frequency locking interval for the ``inverted'' envelope reads
\begin{equation}
\Delta \Omega_{\mathrm{inv}} =\Omega A^2\left[ H_{\mathrm{inv}}(\chi^+_{\mathrm{inv}})-H_{\mathrm{inv}}(\chi^-_{\mathrm{inv}})\right]=\Delta \Omega.
\label{inter_inv}
\end{equation}
In relative units, the power of the ``inverted'' envelope is $\left\langle \psi_{\mathrm{inv}}^2 \right\rangle=1-\left\langle \psi^2 \right\rangle$. Thus, we get an important conclusion: any envelope waveform satisfying $1/2<\left\langle \psi^2 \right\rangle \leq1$ can be considered as an unreasonable stimulation protocol, since its ``inverted'' version gives the same result with the lower cost.

\section{\label{sec_opt} Optimal waveform}

The optimization problem can be formulated as follows. Under the fixed values of the carrier  $\omega$ and modulation $\Omega$ frequencies of the signals and the power of external force $P=T^{-1}K^2\int_0^T  \psi^2(\Omega t)\varphi^2(\omega t) \mathrm{d}t$ one needs to find such $K^*$, $\psi^*(s)$, and $\varphi^*(s)$  that would maximize the frequency locking interval $\Delta \Omega$. Additionally, the external force can not exceed predefined value $I_0\geq|K\psi(\Omega t)\varphi(\omega t)|$. Since both $\psi(s)$ and $\varphi(s)$ at least once during the period hit a value equal to one, the last constrain can be written as $|A|\leq I_0/\omega$.

In the limit of $\omega/\Omega \rightarrow \infty$, value of the function $\psi(\Omega t)$ changes  slightly through the HF force's period~$2\pi/\omega$. Therefore, the power of the external force can be approximated as a product of two factors:
\begin{equation}
P=P_{\psi}P_{\varphi}=\left(\frac{\omega^2}{2\pi}A^2 \int\limits_0^{2\pi}\psi^2(s)\mathrm{d}s\right)\left(\frac{1}{2\pi} \int\limits_0^{2\pi}\varphi^2(s)\mathrm{d}s \right).
\label{power}
\end{equation}
The factor $P_\psi$ depends only on the modulation envelope $\psi(s)$ and amplitude $A$, while factor $P_\varphi$ depends exceptionally on the HF part $\varphi(s)$. Thus, the optimization of the $\psi(s)$ and $\varphi(s)$ waveforms can be accomplished separately.

From Eqs.~\eqref{H_def}, \eqref{inter}, and the definition Eq~\eqref{Phi_def} of the function $\Phi(s)$, one can see that variation of $\varphi(s)$ influences on the frequency locking interval $\Delta \Omega$ only through the multiplier~$\left\langle \Phi^2 \right\rangle$. In Ref.~\cite{Pyragas2015} it was shown that the maximal possible value of the multiplier  $\left\langle \Phi^2 \right\rangle$ is reached with the function $\varphi^{*}(s)=\sin(s+\beta)$ where  $\beta$ is any phase. The power of the optimal HF part would be $P_{\varphi}=1/2$ and $\left\langle \Phi^2 \right\rangle=1/2$.

Further we will consider the problem of envelope $A\psi(s)$ optimization. We seek to maximize $\Delta \Omega$, hence the definition Eq.~(\ref{H_def}) is inserted into Eq.~(\ref{inter}) and from the integrand the Lagrangian of the optimization problem is constructed:
\begin{eqnarray}
& & \mathcal{L}\left(\bar{\phi}^{(+)},\bar{\phi}^{(-)},\psi,\tau\right)= \frac{\Omega A^2}{2\pi}\frac{\left\langle \Phi^2 \right\rangle}{2} \psi^2 \nonumber \\
& &\times \left[\tilde{z}_{\mathrm{eff}}(\chi^+ +\tau)-\tilde{z}_{\mathrm{eff}}(\chi^- +\tau) \right].
\label{lagr}
\end{eqnarray}
The Lagrangian contains the difference of the effective PRCs shifted by the phases $\chi^{\pm}$. It is caused by the fact that the frequency mismatch contains both positive and negative values. In the case of strictly positive (or negative) mismatch, the Lagrangian Eq.~(\ref{lagr}) will have only one effective PRC. The analogous single-sing problem was solved in Ref.~\cite{Pyragas2015}.

We are interested in the time interval $\tau \in [0,2\pi]$. The boundary conditions for the dynamical variables $\bar{\phi}^{(\pm)}(0)=\bar{\phi}^{(\pm)}(2\pi)=\chi^{\mp}$ are satisfied automatically, according to the definition of the points $\chi^{\mp}$. Also the Lagrangian Eq.~(\ref{lagr}) does not depend on $\bar{\phi}^{(\pm)}$, hence these variables can be  ignored.

Considering the requirement for the fixed power $P_{\psi}$, the additional dynamical variable $\kappa(\tau)$ is introduced, which is governed by the differential equation ${\mathrm{d}\kappa/\mathrm{d}\tau=(2\pi)^{-1}A^2\omega^2\psi^2}$. The variable  $\kappa(\tau)$ satisfies the boundary conditions $\kappa(0)=0$, $\kappa(2\pi)=P_{\psi}$. To eliminate the explicit time-dependence, we add the additional dynamical variable $h(\tau)$ governed by the equation $\mathrm{d}h/\mathrm{d}\tau=1$ with the boundary conditions $h(0)=0$ and $h(2\pi)=2\pi$. The Hamiltonian of the system reads
\begin{eqnarray}
& & \mathcal{H}\left(\psi,h,\kappa, p_{h},p_{\kappa}\right) = p_{h}+ p_{\kappa} \frac{\mathrm{d}\kappa}{\mathrm{d}\tau} + \mathcal{L}\left(\psi,h\right) \nonumber \\
& &  =p_{h}+\frac{\Omega A^2 \left\langle \Phi^2 \right\rangle}{4\pi}\psi^2 \nonumber \\
& & \times \left[\tilde{z}_{\mathrm{eff}}(\chi^+ +h)-\tilde{z}_{\mathrm{eff}}(\chi^- +h) +\frac{2\omega^2}{\Omega\left\langle \Phi^2 \right\rangle}p_{\kappa}\right],
\label{ham}
\end{eqnarray}
where $p_{h}$ and $p_{\kappa}$ are the adjoint variables corresponding to $h$ and $\kappa$ respectively. The adjoint equation for the variable $p_{\kappa}$ gives
\begin{equation}
\frac{\mathrm{d}p_{\kappa}}{\mathrm{d}\tau} =-\frac{\partial \mathcal{H}}{\partial \kappa}=0 \Rightarrow p_{\kappa}=\mathrm{const}=-\frac{\Omega\left\langle \Phi^2 \right\rangle}{2\omega^2}C.
\label{kappa_adj}
\end{equation}
According to the Pontriagin's maximum principle~\cite{pontryagin1962}, on the optimal trajectory $\psi^*(\tau)$ the Hamiltonian achieves its maximal possible value. Applying this principle to Hamiltonian defined in Eq.~\eqref{ham}, we see that the optimal envelope is a bang-bang type solution
\begin{equation}
\psi^*(\tau)=
\begin{cases}
1 & \mathrm{when} \: \left[\tilde{z}_{\mathrm{eff}}(\chi^+ +\tau)-\tilde{z}_{\mathrm{eff}}(\chi^- +\tau) -C \right] > 0\\
0 & \mathrm{when} \: \left[\tilde{z}_{\mathrm{eff}}(\chi^+ +\tau)-\tilde{z}_{\mathrm{eff}}(\chi^- +\tau) -C \right] < 0
\end{cases},
\label{psi_opt}
\end{equation}
and the optimal amplitude $A^*$ is the maximal possible value $A^*=I_0/\omega$. The adjoint equation for $p_{h}$ will guarantee a constant value of the Hamiltonian on the optimal trajectory, but it does not give any additional information.

The optimal waveform $\psi^*(\tau)$ is defined up to phase shift. It means that the shifted function $\psi^*(\tau+x)$ also will be an optimal waveform, only the stationary points of Eqs.~\eqref{phase_aver} $\chi^\pm$ will be shifted by the amount $-x$. Thus, without loss of generality, one can take $\chi^-=0$, and then conditions Eqs.~(\ref{psi_opt}) for the optimal envelope simplify to
\begin{equation}
\psi^*(\tau)=
\begin{cases}
1 & \mathrm{when} \: \left[\tilde{z}_{\mathrm{eff}}(\chi^+ +\tau)-\tilde{z}_{\mathrm{eff}}(\tau) -C \right] > 0\\
0 & \mathrm{when} \: \left[\tilde{z}_{\mathrm{eff}}(\chi^+ +\tau)-\tilde{z}_{\mathrm{eff}}(\tau) -C \right] < 0
\end{cases}.
\label{psi_opt1}
\end{equation}
The remaining constants $\chi^+$ and $C$ need to be found using conditions
\begin{subequations}
\label{H_maxmin}
\begin{eqnarray}
\mathrm{max}[H] & = & H(\chi^+),\label{H_max}\\
\mathrm{min}[H] & = & H(0), \label{H_min}
\end{eqnarray}
\end{subequations}
and
\begin{equation}
\left\langle \psi^{*2} \right\rangle=P_{\psi}/I_0^2.
\label{pow_psi}
\end{equation}
Generally, to satisfy these conditions can be a difficult task, but an analytical expression for $\psi^*(\tau)$ can be found for some specific case. In Sec.~\ref{subsec_op_en_an} we consider the function $\tilde{z}_{\mathrm{eff}}(\tau)$ containing a particular symmetry and obtain an analytical expression for the optimal envelope. However, for the case of the FitzHugh-Nagumo oscillator the effective PRC does not have a visible symmetry, thus the optimal envelope is found numerically (see the Sec.~\ref{sec_FHN}).

Noteworthy, the Pontriagin's maximum principle gives only the necessary, but not sufficient conditions. This means that there may be such $\psi$, the constants $\chi^+$ and $C$ that satisfy Eqs.~(\ref{psi_opt1}), (\ref{H_maxmin}), and (\ref{pow_psi}), but the waveform $\psi$ is not optimal. In such cases nonoptimal waveforms need to be filtered out. The FitzHugh-Nagumo oscillator analyzed in Sec.~\ref{sec_FHN} contains such nonoptimal waveforms.

Now we will discuss features of the inverted optimal envelope. Let's say we have constants $\chi^+$ and $C$, which give the optimal waveform $\psi^*(\tau)$. Since $\psi^*$ is equal to $0$ or $1$, the inverted version can be written as ${\psi_{\mathrm{inv}}^*(\tau)=1-\psi^*(\tau)}$. It is also the optimal waveform, but for different power $\left\langle \psi_{\mathrm{inv}}^{*2} \right\rangle=1-\left\langle \psi^{*2} \right\rangle$. The shifted  optimal envelope ${\psi_{\mathrm{inv}}^*(\tau-\chi^+)}$ will satisfy the condition Eqs.~(\ref{psi_opt1}) and (\ref{H_maxmin})  with the constants $\chi_{\mathrm{inv}}^+=2\pi-\chi^+$ and $C_{\mathrm{inv}}=-C$. Since any optimal waveform $\left\langle {\psi^*}^2 \right\rangle \in (1/2,1]$ can be ``inverted'', we will focus only on the waveforms $\left\langle {\psi^*}^2 \right\rangle \in [0,1/2]$.

\subsection{\label{subsec_op_en_an} The case of the symmetric effective PRC}

If the effective PRC has the following symmetry, $\tilde{z}_{\mathrm{eff}}(\tau + \pi)=-\tilde{z}_{\mathrm{eff}}(\tau)$, then according to Eq.~(\ref{H_def}) the function $H(\chi)$ has the same symmetry, $H(\chi+\pi)=-H(\chi)$. Since a satisfaction of the condition Eq.~(\ref{H_min}) implies that the maximum of $H(\chi)$ is at the point $\pi$, the constant $\chi^+=\pi$ for any power values $P_{\psi}/I_0^2$. Then the optimal envelope reads
\begin{equation}
\psi^*(\tau)=\sigma\left(-2\tilde{z}_{\mathrm{eff}}(\tau)-C\right),
\label{psi_opt2}
\end{equation}
where $\sigma(\cdot)$ is the Heaviside step function and $C$ lies in the interval $[-2\Lambda,2\Lambda]$ with $\Lambda=\mathrm{max}[\tilde{z}_{\mathrm{eff}}]$. The power monotonically decreases from $P_{\psi}/I_0^2=1$ at $C=-2\Lambda$ to $P_{\psi}/I_0^2=0$ at $C=2\Lambda$, while the frequency locking interval,
\begin{equation}
\Delta\Omega=-\Omega\left\langle \Phi^2 \right\rangle \left(\frac{I_0}{\omega}\right)^2 \left\langle \sigma(-2\tilde{z}_{\mathrm{eff}}(\tau)-C)\tilde{z}_{\mathrm{eff}}(\tau) \right\rangle,
\label{freq_int}
\end{equation}
monotonically increases from $\Delta\Omega=0$ at the constant $C=-2\Lambda$, until it achieves  maximal value $\Delta\Omega=\Omega\left\langle \Phi^2 \right\rangle  \left(I_0/\omega \right)^2 \left\langle \sigma(\tilde{z}_{\mathrm{eff}}(\tau))\tilde{z}_{\mathrm{eff}}(\tau) \right\rangle$ at $C=0$, and then symmetrically decreases to $\Delta\Omega=0$ at $C=2\Lambda$.
 
Let's take the particular form $\tilde{z}_{\mathrm{eff}}(s)=\sin(s)$. From Eq.~(\ref{psi_opt2}) one can see that the optimal envelope contains only one impulse of a width $P_{\psi}/I_0^2$ with a center placed at $\tau=3\pi/2$. Then the constant $C$ related with the power by $C=2\cos\left(\pi P_{\psi}/I_0^2 \right)$. Another realistic example is the Stuart-Landau oscillator $\dot{w}=(1+i)w-w|w|^2$, where the complex variable $w(t)=x(t)+iy(t)$ incorporates the first $x(t)$ and the second $y(t)$ dynamical variables. The limit cycle $\bm{\xi}(t)=(\cos(t),\sin(t))^T$ has the infinitesimal PRC $\mathbf{z}(\vartheta)=(-\sin(\vartheta),\cos(\vartheta))^T$ and the effective PRC $\tilde{z}_{\mathrm{eff}}(s)=2\sin(2s)$ (if the perturbation is applied to the first dynamical variable). Again, using Eq.~(\ref{psi_opt2}), one can see that the optimal envelope contains two identical impulses with the centers separated from each other by $\pi$.

\section{\label{sec_FHN} Numerical demonstration on FitzHugh-Nagumo neurons}

In this section we will demonstrate the AMHF waveform optimization problem for two-dimensional FitzHugh-Nagumo (FHN) neuron model~\cite{Fitzhugh61,Nag62}. Ensemble of all-to-all mean field coupled FHN neurons is described by the set of the differential equations
\begin{subequations}
\label{modelFHN}
\begin{eqnarray}
\dot{v}_{j} & = & v_j-\frac{v_j^3}{3}-w_j+I_j+I_\text{cp} +K\psi(\Omega t)\varphi(\omega t), \label{FHN_v} \\
\dot{w}_{j} & = & \varepsilon (v_j+ \beta - \gamma w_j), \label{FHN_w}
\end{eqnarray}
\end{subequations}
where $v_j$ denotes the membrane potential and $w_j$  stands for  the recovery variable of $j$-th neuron ($j=1\ldots N$). The direct current $I_j$ defines the spiking frequency of the free neuron. The neurons are coupled through the mean field $I_\text{cp}=g(V-v_j)$, where $g$ is the coupling constant and the variable $V$ stands for the potential generated by the mean field:
\begin{equation}
V=\frac{1}{N}\sum_{i=1}^N v_i.
\end{equation}
The last term in Eq.~\eqref{FHN_v} represents the external AMHF force. We choose the standard values of the parameters $\varepsilon=0.08$, $\beta=0.7$, $\gamma=0.8$.  We select $I_j=0.5$ as a ``central'' neuron, which  natural frequency  ${\Omega_j\approx 0.15918}$~(${T_j\approx 39.47319}$) coincides with the frequency of the envelope $\Omega$. The high-frequency was chosen equal to $\omega=1000\Omega$ and the HF waveform $\varphi(s)=\cos(s)$.
\begin{figure} 
\centering\includegraphics[width=0.99\columnwidth]{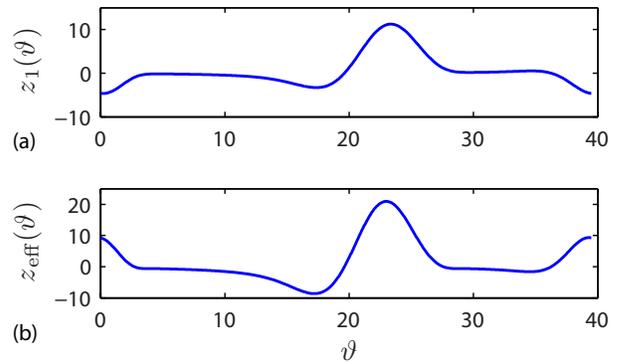}
\caption{\label{fig0} PRC calculated for the FHN neuron with applied direct current $I_j=0.5$. (a) The first component of the infinitesimal PRC and (b) the effective PRC.}
\end{figure}
\begin{figure} 
\centering\includegraphics[width=0.99\columnwidth]{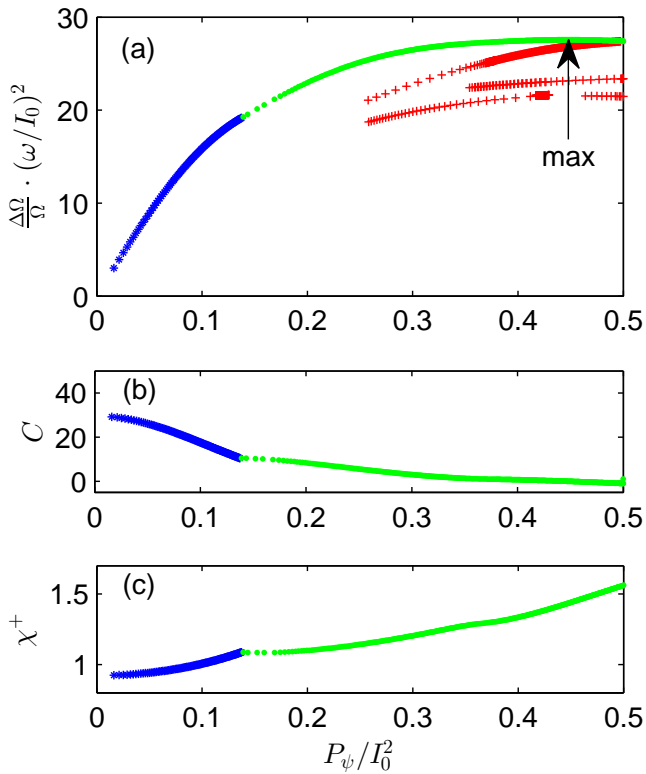}
\caption{\label{fig1_1} (a) The frequency-locking interval estimated from condition Eqs.~(\ref{H_maxmin}). The blue asterisks and green dots represent the optimal envelopes consisting of one and two impulses, respectively, while the red ``plus'' signs represent nonoptimal solutions. The arrow indicates the point where the locking interval achieves maximum. Panels (b) and (c) shows the constants $C$ and $\chi^+$ of the optimal envelope.}
\end{figure}
The uncoupled and unperturbed ``central'' neuron has the infinitesimal and effective PRCs computed numerically and showed in Fig.~\ref{fig0}. One can see that the  $z_1(\vartheta)$ and $z_{\mathrm{eff}}(\vartheta)$ have  similar shapes, since the second derivatives of the vector field is $\partial^2\mathbf{f}(v,w)/\partial v^2=(-2v,0)^T$ and only the first component of the infinitesimal PRC has influence on the effective PRC.

Further, we use this numerically calculated effective PRC to construct $\psi(\tau)$ according to Eq.~(\ref{psi_opt1}) condition. We scan through all possible constants $\chi^+$ and $C$, and find such pairs, which gives the envelopes satisfying the condition Eqs.~(\ref{H_maxmin}). In Fig.~\ref{fig1_1}, blue asterisks and green dots show the frequency locking interval and corresponding constants $\chi^+$, $C$ dependence on the relative power of the optimal envelope. The red ``plus'' signs in Fig.~\ref{fig1_1}(a) mark the nonoptimal envelopes, which also satisfy the condition Eqs.~(\ref{H_maxmin}), but gives lower locking interval and therefore are not shown in Figs.~\ref{fig1_1}(b) and \ref{fig1_1}(c). Such nonoptimal solutions appears due to the fact that the Pontriagin's maximum principle is necessary but not sufficient. For small power values $P_\psi/I_0^2<0.14$ the optimal envelope contains one impulse (blue/dark color), but for larger powers the second impulse in the waveform $\psi^*$ appears (green/light color). We limit ourselves by $P_\psi/I_0^2\in [0,0.5]$, since using ``inversion'' symmetry all results for $P_\psi/I_0^2\in [0.5,1]$ can be recovered from the Fig.~\ref{fig1_1}. The frequency-locking interval [Fig.~\ref{fig1_1}(a)] achieves maximal value at $\left\langle \psi^{*2} \right\rangle \approx 0.45$, so all envelope waveforms with the power higher than $0.45$ can be considered as unreasonable stimulation protocols in that sense, that the same locking interval can be achieved with lower power costs.
\begin{figure} 
\centering\includegraphics[width=0.99\columnwidth]{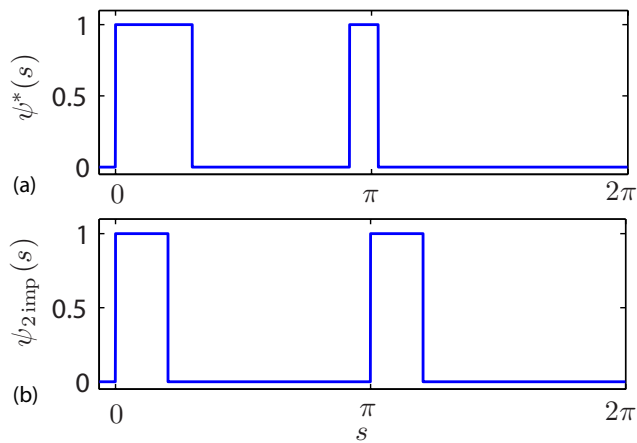}
\caption{\label{fig2}(a) The optimal envelope with the power $P_{\psi}/I_0^2\approx 0.21$. (b) The non-optimal envelope of two impulses with the same power as in the case (a).}
\end{figure}

\begin{figure} 
\centering\includegraphics[width=0.99\columnwidth]{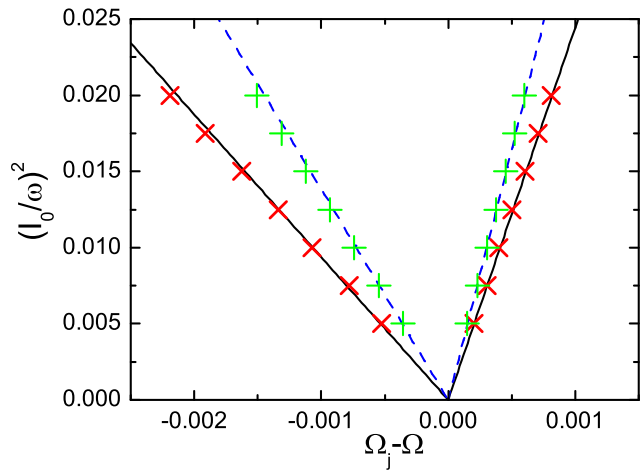}
\caption{\label{fig3} The Arnold tongues for the perturbed FHN neuron by the AMHF signals with the envelopes presented in Fig.~\ref{fig2}. The HF $\omega=1000\Omega$ remains fixed, while $I_0$ and $\Omega_j$ vary. Continuous and dashed lines represent the minimal and maximal frequencies calculated from the phase reduction theory for the optimal [Fig.~\ref{fig2}(a)] and nonoptimal [Fig.~\ref{fig2}(b)] envelopes, respectively. These lines are compared with synchronization regions estimated from direct integration of the FHN neuron, marked by red crosses for optimal and green plus signs for non-optimal envelopes.}
\end{figure}

For the numerical demonstrations we fix the power of the envelope $\left\langle\psi^{*2}\right\rangle=P_{\Omega}/I_0^2 \approx 0.21$. The optimal envelope waveform is depicted in Fig.~\ref{fig2}(a) together with non-optimal envelope waveform $\psi_{2\,\mathrm{imp}}$ containing two impulses of equal width (see Fig.~\ref{fig2}(b)), which gives the same power. For both waveforms we calculate the Arnold tongues showed in Fig.~\ref{fig3}. The analytical values (depicted by the straight lines) are calculated from Eq.~(\ref{bound_freq}) while the numerical results (depicted by the symbols) are obtained by integrating the single FHN neuron [$g=0$ in Eq.~(\ref{modelFHN})] with different direct current $I_j$ for different frequency mismatch $\Omega_j-\Omega$. As predicted, the optimal envelope $\psi^*$ gives the higher frequency-locking interval than the nonoptimal envelope $\psi_{2\,\mathrm{imp}}$.
\begin{figure} 
\centering\includegraphics[width=0.99\columnwidth]{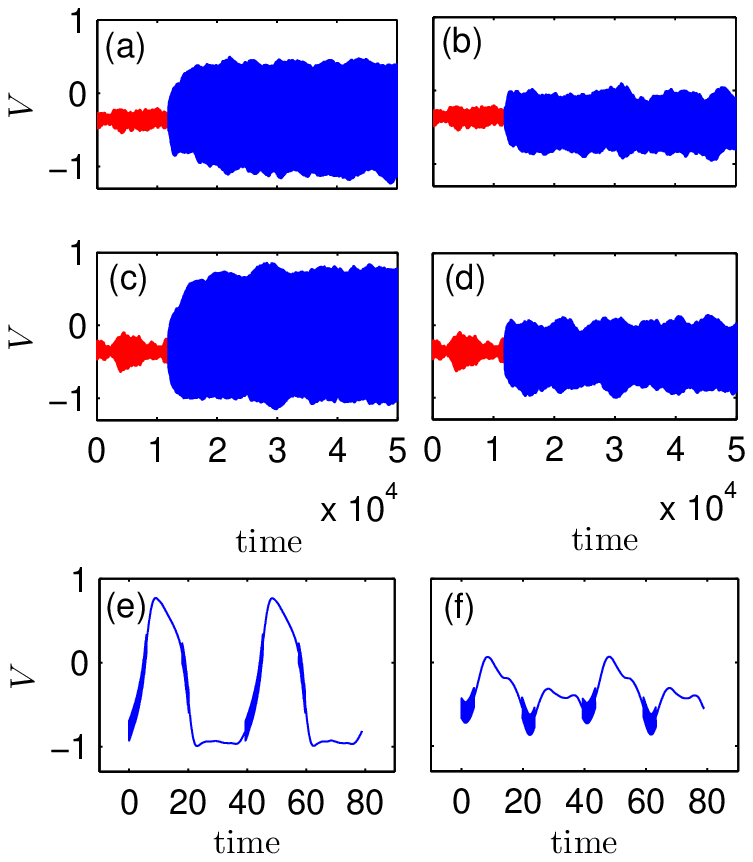}
\caption{\label{fig4} An example of the mean-field dynamics of $N=300$ uncoupled $g=0$, panels (a) and (b), and coupled $g=0.001$, panels (c)--(f), FHN neurons. In panels (a)--(d) oscillators evolves freely for the time $t=300 T$ drawn by red line. From $t=300T$, (a) and (c) systems are affected by the AMHF signal with the optimal envelope, while panels (b) and (d) represent nonoptimal [Fig.~\ref{fig2}(b)] stimulation. The stimulation amplitude is $I_0=19.5$. Panels (e) and (f) show dynamics distinguished from panels (c) and (d), respectively. The direct currents $I_j$ applied to neurons are selected in such a way that the natural frequencies would be uniformly distributed in the interval $\Omega_j \in [-0.002,0.001]+\Omega$.}
\end{figure}

In order to demonstrate an advantage of the optimal over nonoptimal envelope in application to ensemble of the oscillators, we numerically simulate $N=300$  uncoupled and coupled FHN neurons.  Figures~\ref{fig4}(a) and \ref{fig4}(b) show the dynamic of the mean field of free (red/light color, till $t=300T\approx1.2 \cdot 10^4 $) and stimulated (blue/dark color, for $t>300T$) system Eq.~\eqref{modelFHN}. In Figs.~\ref{fig4}(a), \ref{fig4}(c), and \ref{fig4}(e), the system undergoes an optimal control force with envelope $\psi^*(\Omega t)$ showed in Fig.~\ref{fig2}(a), while in Figs.~\ref{fig4}(b), \ref{fig4}(d), and \ref{fig4}(f) system is controlled by nonoptimal force with the envelope of two pulses showed in Fig.~\ref{fig2}(b). Comparing the mean fields we see that in the case of the optimal envelope it has a larger amplitude of oscillations than in the nonoptimal case. This means that in the first case more neurons are entrained by the external force. Note that our developed theory gives the optimal waveform for uncoupled oscillators only. However, one can see that even for the weakly coupled case the optimal waveform [see Fig.~\ref{fig4}(c)] gives higher mean field compared to nonoptimal waveform [see Fig.~\ref{fig4}(d)]. In the enlarged graphics, one can see that the shape of the mean field Fig.~\ref{fig4}(e) recalls single neuron dynamics, while in Fig.~\ref{fig4}(f) it is quite different.

\section{\label{sec_disc} Conclusions}

In conclusion, we have developed the algorithm for obtaining the optimal bounded envelope waveform for the oscillators affected by the strong amplitude-modulated high-frequency external force to ensure the maximal frequency locking interval. Using the Pontriagin's maximum principle, we obtain that the optimal waveform is a bang-bang-type solution. The conditions for an estimation of the particular time moments where the external force must be turned on and off were derived. These conditions depend on a shape of the PRC. For small power values the optimal envelope usually has only one impulse of stimulation while for the higher power it becomes more complex. Additionally, we have shown that any envelope waveform can be ``inverted'' and still will have the same frequency locking interval. Our theory is illustrated numerically on the FitzHugh-Nagumo oscillators. Although, the theory is derived for uncoupled oscillators, the generalization for the case of coupled oscillators may be achieved by utilization of the collective phase response function~\cite{kawam08}. Our findings are relevant to the design of mild neurostimulation protocols for treatment of neurological diseases.

\bibliographystyle{apsrev4-1}
\bibliography{references}

\end{document}